\documentclass[draft,jgrga]{agutex}

\usepackage{pdflscape}
  \usepackage[dvips]{graphicx}
 \setkeys{Gin}{draft=false}
\authorrunninghead{LIU and QIN }
\titlerunninghead{Improvements on STOA}
\authoraddr{H.-L. Liu,
State Key Laboratory of Space Weather,
Center for Space Science and Applied Reaserch, Chinese Academy of Sciences,
P.O. Box 8701, Beijing 100190, China.
(hlliu@spaceweather.ac.cn)}
\authoraddr{G. Qin,
State Key Laboratory of Space Weather,
Center for Space Science and Applied Reaserch, Chinese Academy of Sciences,
P.O. Box 8701, Beijing 100190, China.
(gqin@spaceweather.ac.cn)}

\begin{document}
\bibliographystyle{agu08}

%% ------------------------------------------------------------------------ %%
%
%  TITLE
%
%% ------------------------------------------------------------------------ %%

\title{Improvements of the shock arrival times at the Earth model STOA}
% New Methods of Predicting the Shock Arrival Times (SATs) at the Earth 
%STOA$^\prime$, STOASSN and STOAAD }
%
% e.g., \title{Terrestrial Ring Current:
% Origin, Formation and Decay $\alpha\beta\Gamma\Delta$}
% You may use \\ to break the title over several lines.

%% ------------------------------------------------------------------------ %%
%
%  AUTHORS AND AFFILIATIONS - 2 methods
%
%% ------------------------------------------------------------------------ %%

% Method 1 
% For three or fewer author/affiliation blocks, use \author{} and \affil{}

\author{H.-L. Liu and G. Qin   }
\affil{State Key Laboratory of Space Weather,
Center for Space Science and Applied Research, Chinese Academy of Sciences,
P.O. Box 8701, Beijing 100190, China}

% ---------------
% Method 2 
% For more than three author/affiliation blocks,
% use \author{\altaffilmark{}} and \altaffiltext{}
% \altaffilmark will produce footnote;
% matching altaffiltext will appear at bottom of page.
% May use \\ to start a new line.

% \authors{R. C. Bales, \altaffilmark{1}
% E. Mosley-Thompson, \altaffilmark{2}
% R. Williams, \altaffilmark{3}
% and J. R. McConnell\altaffilmark{4}}

% \altaffiltext{1}
% {Department of Hydrology and Water Resources, University of Arizona,
% Tucson, Arizona, USA.}
%
% \altaffiltext{2}{Department of Geography, Ohio State University,
% Columbus, Ohio, USA.}
%
% \altaffiltext{3}{Department of Space Sciences, University of Michigan,
% Ann Arbor, Michigan, USA.}
%
% \altaffiltext{4}{Desert Research Institute, Division of Hydrologic Sciences,
% Reno, Nevada, USA.}

%% ------------------------------------------------------------------------ %%
%
%  ABSTRACT
%
%% ------------------------------------------------------------------------ %%

% >> Do NOT include any \begin...\end commands within
% >> the body of the abstract.

\begin{abstract}
Prediction of the shocks' arrival times (SATs) at the Earth is very important for space weather forecast. There is a well-known SAT model, STOA, which is widely used in the space weather forecast. However, the shock transit time from STOA model usually has a relative large error compared to the real measurements. In addition, STOA tends to yield too much `yes' prediction, which causes a large number of false alarms. Therefore, in this work, we work on the modification of STOA model. First, we  give a new method to calculate the shock transit time by modifying the
way to use the solar wind speed in STOA model. Second, we  develop new criteria for deciding whether the shock will arrive at the Earth with the help of  the sunspot numbers and the angle distances of the flare events. It is shown that our work can improve the SATs prediction significantly, especially the prediction of flare events without shocks arriving at the Earth.
\end{abstract}

%% ------------------------------------------------------------------------ %%
%
%  BEGIN ARTICLE
%
%% ------------------------------------------------------------------------ %%

% The body of the article must start with a \begin{article} command
%
% \end{article} must follow the references section, before the figures
%  and tables.

\begin{article}

%% ------------------------------------------------------------------------ %%
%
%  Introduction
%
%% ------------------------------------------------------------------------ %%
\section {Introduction}
Interplanetary shocks are among the products of eruptive solar events, and they can 
also accelerate energetic particles which influence the geo-space environment 
seriously. So predicting of the shock arrival times (SATs) at Earth is a necessary 
step for space weather forecast system. Coronal shock waves are counterparts of
 the interplanetary shocks in corona. The coronal shock waves' origin is still 
questionable, but generally two eruptive phenomena from the Sun  are thought to be 
responsible: flares and coronal mass ejections (CMEs) \citep{VrsankACliver08}.  So  
most prediction models take the observation data of flares or CMEs as their inputs. 
Although CMEs are usually associated with flaress \citep{Dryer96}, in this paper we 
mainly focus on the shocks related with solar flare eruption events.
%Generally, the interplanetary shocks are 
%considered to be caused by flares or coronal mass ejections (CMEs) 
%\citep{VrsankACliver08}.
%So SAT prediction models are divided into two categories, models to predict
%flare-originated shock events and CME-originated ones. 
%Here, we mainly focus on the shocks caused by solar flare events.
 
Among many models built to predict the  eruption-driven shock events, there are   
three famous physics-based  eruption-driven shock SAT prediction models, 
the Shock Time of Arrival (STOA) model \citep{DryerASmart84,
SmartAShea85}, the Interplanetary Shock 
%SmartEA84,SmartEA86,,LewisADryer87
Propagation Model (ISPM) \citep{SmithADryer90}, and the Hakamada-Akasofu-Fry Version
 2 (HAFv.2) Model \citep{FryEA01}, which are used  widely and referred as 3PMs 
hereafter.  Afterwards, many 
other SAT prediction models are developed \citep[e.g.,][]{FengAZhao06, FengEA09a, FengEA09b,
QinEA09jgr, LiuAQin12}.  For the prediction of CMEs-related shocks, please refer to \citet{ZhaoADryer14}.

It is shown that 3PMs use the same observation data as inputs. As numerical 
simulation models considering more physics mechanisms, ISPM and HAFv.2 are more
complicated than STOA, which is an  analytical model. Especially, the HAFv.2 model is 
rather complicated, which simulates the magnetic fields in solar wind with the 
effects of the transitions of shocks.  ISPM is based on 2.5 MHD simulation, and HAFv.2 is calibrated with 1D and 2D MHD simulation \citep{SunEA85}.  
So, STOA is relatively easier 
to operate and optimize, but its performance is not inferior to ISPM and HAFv.2 models \citep{FryEA03, McKennaLawlorEA06}. 
In practice, the STOA model is used 
most extensively among 3PMs. %And the work done in this paper is based on it.  

%Transit time of shock is the most important output parameter of a SATs prediction model. The three famous model mentioned above (STOA, ISPM and HAFv.2) gave quite similar root mean square errors for the $\Delta T$ (defined as the observed transit time minus the predicted transit time), even though HAFv.2 introduced the interaction between shock and the interplanetary medium \citep{FryEA03}. 
 
%For example, the initial shock speeds given by \citet{FryEA03,McKennaLawlorEA06,SmithEA09a} are much smaller than given by \citet{ReinerEA07}. 

Transit time of the shock is an important output for SAT prediction models.  
STOA yields $12$ hours root mean square (RMS) error for $\Delta T$ (difference 
between predicted transit time $T_{pre}$ and observed transit time $T_{obs}$). 
Besides other effects, the inaccuracy of the input parameters  contribute to the 
forecast errors of SAT prediction models.
 \citet{ZhaoAFeng14} built a new prediction model SPM2 by adjusting the input 
parameters of SPM model, and SPM2 performed better than SPM. 
In STOA model, shock is assumed to ride over an  isotropic background solar wind
plasma flow, so the speed ($V_{sw}$) of the flow is one of the input parameters of STOA. However,
the shock goes through background solar wind flow with varying $V_{sw}$ during its 
propagation from the Sun to the observer, and no observation techniques at present 
could provide $V_{sw}$ accurately over the whole path route of the shock.  
Furthermore, the $1$ AU solar wind velocity at the time of the parent flare event's
 eruption is used as the speed of the flow in STOA.  It is possible to 
improve the prediction of STOA by providing more reasonable value of $V_{sw}$. 

Not all the coronal shocks would arrive at Earth, eg., some of them may decay to 
MHD waves which are relatively harmless to the electronic instruments. So a model to
predict SATs should use some criterion to predict whether the shock will arrive at 
Earth before predicting the shock's transit time. 
STOA uses a simplified magnetoacoustic Mach number $M_{\alpha}$ at $1$ AU to 
represent the strength of the shock.
%, which is defined as 
%\begin{linenomath*}
%\begin{equation}
%M_a=\frac{V_s}{\sqrt{a^2+V_{\alpha}^2}},
%\label{equ:Mstoa}
%\end{equation}
%\end{linenomath*}
%here $V_s$ is the velocity of the shock front in a solar wind coordinate system, $a$ is the sound speed and $V_{\alpha}$ is Alfven speed. To get  $a$ and $V_{\alpha}$, typical solar wind parameters are used.
If $M_{\alpha}$  is  greater than $1.0$ the model will provide the `yes' prediction 
which means there will be a shock observed at Earth, otherwise the model provides 
the `no' prediction. 
In practice, false `no' prediction may bring damage to satellites in space and some 
electronic instruments on the ground, 
and false `yes' prediction may lead to extra expenses on operation of the 
instruments and discontinuities of science observations. It is shown that STOA 
model tends to provide too much `yes' prediction \citep{McKennaLawlorEA06}. 

Some research about the criterion for  SATs prediction models had been done. Solar 
flare eruption events not only produce coronal shocks but also emit high energy particles. 
\citet{QinEA09jgr} used the $38-53$ KeV electrons observed by EPAM/ACE at $1$ AU 
to help predicting whether the shock will arrive at Earth. New method given by 
\citet{QinEA09jgr} improved the prediction of the SATs significantly. 
\citet{LiuAQin12} developed a new SATs prediction method (STOASF) with the help of 
energy released by flare in soft X-ray. 
%In the method STOASF, a new criterion is created by using a simplified form of $E_x$.
%\begin{linenomath*}
%\begin{equation}
%E_x^\prime=(f-f_0)(\tau-\tau_0),
%\label{equ:liuEx}
%\end{equation}
%\end{linenomath*}
%here $f$ is the peak intensity of the soft X-ray related with the flare event,  $\tau$ is the duration of the flare event. $f_0$ and $\tau_0$ are constants. 
It is shown that the new method STOASF performs much better than STOA on those 
events without shock arrivals at Earth \citep{LiuAQin12}.

We build four new SATs prediction methods in this paper. We first
 describe  the data and events used in section 2. In Section 3, we introduce some
new methods to predict SATs.
In section 4, we compare the performances of different prediction methods. We discuss 
and summarize our results in section 5.

%%------------------------------------------------------------------------ 
\section{Data and Events Selection}
The parent solar flare eruption events and their corresponding shocks observations at $1$ AU used 
in our work are from the combination of the lists in \citet{FryEA03}, \citet{McKennaLawlorEA06} and 
\citet{SmithEA09b}. There are $625$ flare events altogether in the three papers, which cover the period of the whole Solar Cycle 23 from February 1997 to December 2006.  The whole data set used in our work is from real time experience unlike the other models regardless of their scientific sophistication. 
STOA only consider the flare energy released during the solar eruption events, so
it made made prediction of the SATs without considering CMEs data\citep{SmithEA00}. 
 \citet{SmithEA09b}  matched CMEs observation data for the event set used in their work and tested the performance of the HAFv.2 model. No CMEs data were included in \citet{FryEA03} and \citet{McKennaLawlorEA06}. In our work, we still don't take account of CMEs. Start times of the metric type II coronal events are taken as the begin times of these parent solar flare eruption events, and the duration are derived from the GOES X-ray flux. Optical flare observation data provides the location of the parent events. The corresponding shocks data at L1 are from ACE, SOHO or WIND. Matching of the parent eruption events and shocks at Earth will be a little different for each model. Here we use the matches that are most favourable for STOA.

 We only use $582$ events in our study, which means $43$ events are excluded from the list mentioned in the last paragraph.  \citet{QinEA09jgr} excluded $3$ events because of the 
lack of  $0.038-0.053$ MeV energetic electrons data observed by ACE/EPAM. 
And another $37$ events with soft X-ray intensity data unavailable were further 
removed from the work of \citet{LiuAQin12}. 
In this paper we remove additional $3$ events with abnormal observed transit 
times, which leaves $582$ events finally. Hereafter, we denote the $582$ events
used in this paper as $E_{582}$.
The fearless forecast numbers of the $43$ events removed from the combination of
the lists in \citet{FryEA03}, 
\citet{McKennaLawlorEA06} and \citet{SmithEA09b} are shown in table \ref{tbl:ffn}. 
Out of the $582$ flare events we use here, $225$ are accompanied with shocks at 
Earth, which are named as `with shock' (WS) events, and the rest $357$ events 
without shock at Earth are named as `without shock' (WOS) events.

%---------------------------------------------------------------------------
\section{New Prediction Methods}
In the following, we describe some new prediction methods of SATs. The detailed 
performances check of the methods will be shown in Section 4.
\subsection{Replacing Solar Wind Speed with a Typical Constant $V_c$, a New 
Prediction Method STOA$^\prime$} 
In STOA model, the background solar wind speed ($V_{sw}$) over the whole path 
route of 
the shock is very important for the calculation of the transit time. 
But there is no way to observe such quantity accurately. So STOA used $1$ AU 
solar wind speed at the time of flare events' eruption instead. This process may
yield large errors because of the spacial and temporal perturbations of the solar 
wind. 

%Here we offered another way to use the background solar wind speed in 
%the SATs prediction model,  and a modification version of STOA will be given.  
Therefore, we use a constant $V_c$ to represent the typical value of
solar wind speed over the shock journey for simplicity purpose, and a modification
of the STOA model is obtained. We average solar wind speed over the $E_{582}$ as 
$\overline V_{sw}=455$ km/s. So we can set $V_c=455$ km/s and the modified STOA 
model is denoted as STOA$^\prime$. It should be noted that the input parameters of 
the STOA$^\prime$ are the same as STOA except that $V_{sw}$ is replaced by 
$V_c=455$ km/s in STOA$^\prime$. Furthermore, in STOA$^\prime$, we still use 
$M_{\alpha}$ defined by STOA to measure the strength of the shock at $1$ AU, and 
only when $M_\alpha > 1.0$ would the shock arrive at the Earth. Since $V_{sw}$ is 
not needed to calculate $M_\alpha$ in STOA, the criterion of STOA$^\prime$ is the 
same as that of STOA.

%==============================================================
   
\subsection{Using Sunspot Number (SSN) to Help Predicting the SATs 
at Earth, a New Prediction Method STOASSN}
When using the $E_{582}$ to test performance of STOA's 
criterion, we found that STOA model tends to yield too much `yes'
prediction. There are $225$ of the $582$ flare events followed by 
shocks at Earth,  but STOA model yields $438$ `yes' prediction,  
which may imply that STOA underestimates the decay of shocks by 
various structures in solar wind while propagating through the 
interplanetary space. In addition, there are more structures 
with stronger solar activity, which can be indicated by the 
sunspot number (SSN). Therefore, we introduce a new parameter 
$M_{\alpha}^{SSN}$ to modify $M_{\alpha}$ of STOA with the help of SSN,
%\begin{linenomath*}
\begin{equation}
M_{\alpha}^{SSN}=\frac{M_{\alpha}}{k_{SSN}\overline{SSN}+b_{SSN}}
\label{equ:Mssn}
\end{equation}
%\end{linenomath*}
where $k_{SSN}$ and $b_{SSN}$ are constants, and $\overline{SSN}$ 
is the average of daily sunspot number during a time period before 
the flare event. To get a reasonable $\overline{SSN}$, 
the time period for average of daily sunspot numbers is very 
important.  A too long period would introduce the effects of 
other solar eruption events which are not related to the shock events. Here we use  $10$ day before the flare event as the time period to get the $\overline{SSN}$. 
We can adjust the values of $k_{SSN}$,  $b_{SSN}$ to get 
different models. Using the $E_{582}$ we can 
check the performances of the models. Hereafter we use $k_{SSN}=0.006$ and 
$b_{SSN}=1.1$ which are the best parameters we can get so far. 
Note that the daily sunspot number data are from 
http://sidc.oma.be/html/dailyssn.html. 

By modifying the $M_\alpha$ of STOA as $M_\alpha^{SSN}$ we can get a new SATs
prediction method, STOASSN. So the criterion of STOASSN is different than STOA, but
the transit time of STOASSN is the same as that of STOA.
%So when a flare event occurs, one can predict if the shock 
%will arrive at Earth with $M_{\alpha}^{SSN}$. We still use 
%$1.0$ as the threshold for $M_{\alpha}^{SSN}$.
%If $M_{\alpha}^{SSN} \geq 1.0$, our method will give a `yes' 
%prediction, which means the shock will arrive at the Earth.
%If $M_{\alpha}^{SSN} < 1.0$, a `no' prediction will be given, 
%i.e. the shock will not arrive at the Earth. If a `yes ' prediction 
%is given, the transit time of the shock will be calculated using 
%STOA model. We name the new method as STOASSN. Performance of 
%the new method STOASSN will be given in section 5.
%-----------------------------------------------
\subsection{A New SATs prediction model, STOASSN$^\prime$, Which Combines STOA$^\prime$ and STOASSN }
 Here we introduce a new model STOASSN$^\prime$ by combining two new models, 
STOA$^\prime$ and STOASSN. In STOASSN$^\prime$ model, the background solar wind 
speed $V_{sw}$ is replaced by typical constant $V_{c}=455$ km/s, and meanwhile the 
$M_{\alpha}^{SSN}$ is used as the criterion to decide whether the shock will arrive 
at the Earth. If $M_{\alpha}^{SSN}>1.0$, then a `yes' prediction will be given, 
which means a shock is going to impact the Earth, while if 
$M_{\alpha}^{SSN} \le 1.0$ then no shock will arrive at the Earth.
%---------------------------------------------
\subsection{Using the  Flare Events' Angle Distance to Help Predicting the SATs at Earth, a New Prediction Method STOAAD}
The  angle distance $\phi$ between the shock's nose propagation 
 direction and the Sun-Earth line is also important for SATs prediction. 
 It is known that shock nose is the strongest part of  shock front, 
 and the shock strength decreases by increasing solid angle 
 distance from shock nose. In STOA model,  shock nose is assumed to be in the
direction of the parent flare event. Therefore, the angle distance $\phi$ can be
 calculated from the equation $\cos(\phi)=\cos(\theta)\cos(\varphi)$, where  
 $\theta$ and $\varphi$ are  the central meridian
  diastance and latitude of the solar flare events, 
respectively. Larger angle distance ($\phi$) means that the observer is far away
from the shock nose direction and the  chance to observe a shock is smaller.  
In addition, flare events with higher energy may drive a stronger coronal shock
which is not easy to decay to MHD wave.
\citet{LiuAQin12} used the effects of energy  released in soft X-ray during the flare
events, $E_x^\prime$, to help to decide if shock would arrive at Earth.

Therefore, we build a new criterion by combining the effects of energy $E_x^\prime$ 
and the angle distance $\phi$. Note that in \citet{LiuAQin12} the energy 
$E_x^\prime$ could be negative. Here in order to keep $E_x^{\prime}$ positive for
 convenience, we redefine the $E_x^{\prime}$ as, 
%\begin{linenomath*}
\begin{equation}
E_x^\phi=f\tau,
\label{equ:ExPre2}
\end{equation}
%\end{linenomath*}
where $f$ is the peak intensity of soft X-ray during the 
flare event, and $\tau$ is the duration time of the event. 

\begin{figure}
\noindent\includegraphics[width=25pc]{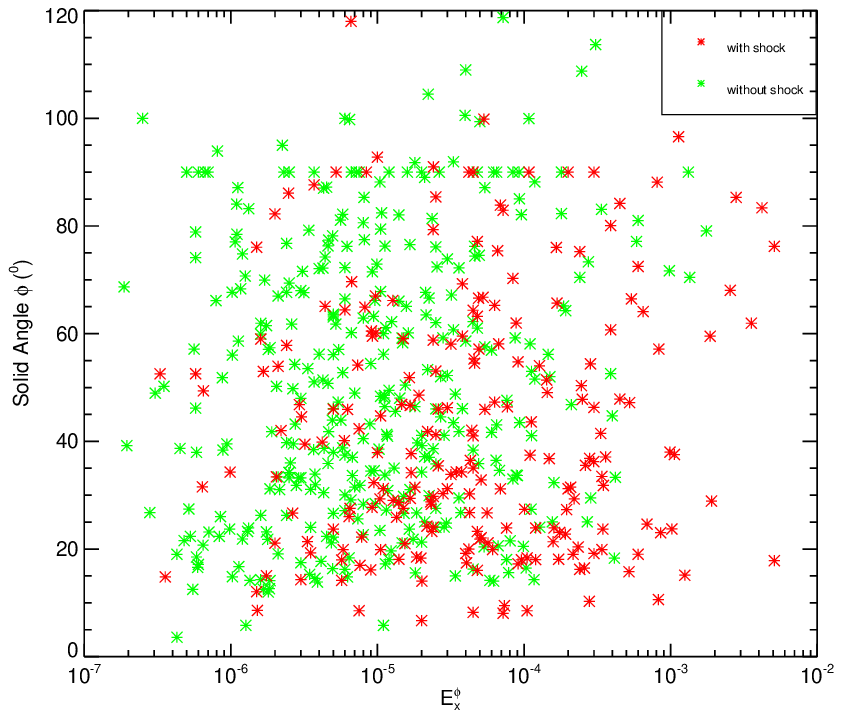}
\caption{Distribution of $582$ events with the energy $E_x^{\phi}$ and the angle distance $\phi$
\label{fig:saftau}}
\end{figure}

Figure \ref{fig:saftau} shows $E_x^\phi$ versus $\phi$   of 
the $E_{582}$. Here we use stars with different colors to 
represent the two type events, red for WS events (the flare events 
that are accompanied by shocks at Earth) and green for WOS events (the flare events
 that are not accompanied by shocks at Earth).  It is shown that 
the WOS events tend to locate in the top left corner of the figure, 
and the WS events are more likely to locate in the bottom right corner. 
%Large angle distance implies the observer is towarding 
%flank of the shock, and flank decay more quickly than nose. 
%Therefore, the larger the angle distance, the lower chance 
%for the observer to detect a shock at Earth.  From figure 
%\ref{fig:saftau}, we also can see that the WS events possess 
%higher $E_x^{\phi}$ than WOS events. This is because  flare 
%event with higher energy may drive a stronger coronal 
%which is not easy to decay to MHD wave.

Based on the above analysis, we combine the effects of $E_x^\phi$ and the  
angle distance $\phi$ to introduce a new criterion for predicting the SATs,
%\begin{linenomath*}
\begin{equation}
C_x^\phi=-\lg\left(\frac{E_x^\phi}{E_{x0}^\phi}\right)
\left(\frac{\phi}{10}\right)^b,
\label{equ:C}
\end{equation}
%\end{linenomath*}
here $E_{x0}^\phi=1$ Whr/m$^2$ is a constant and $E_x^\phi$ is always smaller than
$E_{x0}^\phi$.
Equation \ref{equ:C} shows a negative relationship 
between $E_x^\phi$ and $C_x^\phi$ and a positive 
relationship between $\phi$ and $C_x^\phi$. 
It is possible the two kinds of events, WS events and WOS events, can be separated 
according to the value of $C_x^\phi$. We tried different values of $b$ in 
Equation \ref{equ:C} and found that $b=0.1$ could make 
the best results for the $E_{582}$.  

\begin{figure}
\noindent\includegraphics[width=25pc]{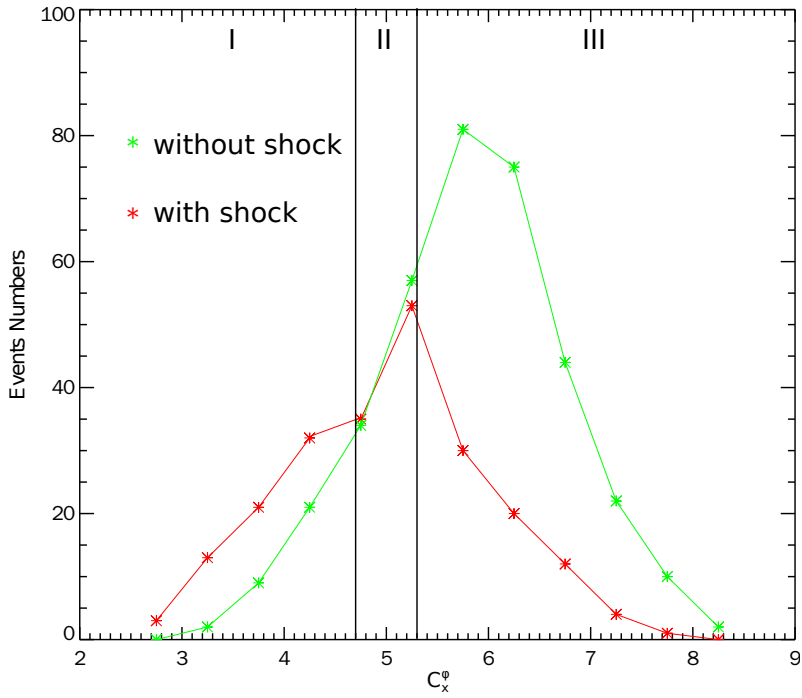}
\caption{Distribution of events' numbers  with $C_x^{\phi}$
\label{fig:Cx}}
\end{figure}

Figure \ref{fig:Cx} shows the number of flare events in different intervals of 
$C_x^\phi$ with the same range $0.5$. 
%The range of $C_x^\phi$ for each data point is $0.5$. 
Red and green symbols  represent WS and WOS events, respectively. The two vertical 
lines, $C_x^\phi=4.7$ and $C_x^\phi=5.3$, divide the x-axis into three regions. 
 It is shown that there are more WS events than WOS events with $C_x^\phi < 4.7$
(Region I), 
which means that shock events in such condition are more likely to arrive at Earth, 
but in contrast most flare events with $C_x^\phi > 5.3$ (Region III) 
are not accompanied by 
shocks at Earth. In addition, with $4.7\le C_x^\phi \le 5.3$ (Region II), 
the ratio of WS events number and WOS events number is close to $1$, which 
implies the probability for a shock to be detected at Earth 
is $50\%$. We take different methods to predict whether the shock 
will arrive at Earth in the three $C_x^\phi$ intervals.  
In STOA model, only the shocks with $M_\alpha > 1.0 $ will  
arrive at the Earth. We lower the threshold of $M_\alpha$ in 
Region I. In this region, if $M_\alpha >0.8$, a `yes' prediction will be provided. 
However, a larger threshold 
for $M_\alpha$, $1.5$, is used in region III, which means that only if
$M_\alpha > 1.5$ a `yes' prediction will be provided. In region II, furthermore,
we adopt the STOASEP model because its performance on WS events 
and WOS events are equal. Here, the new method developed is named STOAAD.

%--------------------------------------------------------------
%\section{Using A Different Initial Energy of Flare Events Given by ISPM Model to Help Predicting SATs at Earth}

% \citet{SmithADryer95} used the net energy of the flare events to calculate the transit time of the shocks. In this section, we develop a new criterion with the help of $E_ispm$. Figure \ref{fig:Eispm} demonstrates the events number versus $E_ispm$ 
 
%---------------------------------------------------------------------------

%----------------------------------------------------------
\section{Performances of the New Methods}
Some variables  are used 
to test the performances of SATs prediction models. The four parameters, hits (h), 
misses (m), false alarms (fa), and correct nulls (cn) are used to express the 
success or failure of the forecasts. In addition, success rate (sr), is an important parameter for the evaluation of SATs 
prediction models. But a prediction model with high value of sr does not 
necessarily guarantee it a good model. Standard 
meteorological skills are introduced to help evaluating the 
performances of the prediction models. Definitions or calculations of these variables are listed as follows,
 \begin{itemize}
 \item Hit (h),\quad shock is predicted and observed at the Earth within $\pm 24$ hours
 \item Miss (m),\quad shock is observed but not predicted within $1 \sim 5$ days of the solar flare event or shock is predicted and observed but not $\pm 24$ hours
 \item False Alarm (fa),\quad shock is predicted but not observed within $\pm 24$ hours
 \item Correct Null (cn),\quad shock is not predicted and no one is observed within $1 \sim 5$ days of the solar flare event
 \item Success Rate (sr),\quad (h+cn)/(h+m+fa+cn)
 \item Probability of detection,\quad yes (PODy),\quad h/(h+m)
 \item Probability of detection,\quad no (PODn),\quad cn/(fa+cn)
 \item False alarm ratio (FAR),\quad fa/(h+fa)
 \item Bias,\quad (h+fa)/(h+m)
 \item Critical success index (CSI),\quad h/(h+fa+m)
 \item True skill score (TSS),\quad PODy+PODn-1
 \item Heidke skill score (HSS),\quad (h+cn-C1)/(N-C1)
 \item Gilbert skill score (GSS),\quad (h-C2)/(h+fa+m-C2)
 \end{itemize}
\noindent  where N (N=h+m+fa+cn) is the total number of  events used in our study, C2=(h+m)(h+fa)/(N), and C1=C2+(fa+cn)(m+cn)/N.  The $E_{582}$ and their corresponding observations at 
Earth  described in section 2 are used to test the performances of STOA and other four new models,
STOA$^\prime$, STOASSN, STOASSN$^\prime$, and STOAAD. 
 
It can be seen from table \ref{tbl:4result} that STOA$^\prime$ 
and STOA nearly yield the same results, but STOA$^\prime$ predicts 
one more WS event correctly than STOA does. 
 The new models STOASSN, STOASSN$^\prime$ and STOASD miss more shock events than STOA does, but 
they correctly predict more WOS events than STOA does. So the 
number of correct nulls yielded by the three of them  is much larger than that 
by STOA and STOA$^\prime$. There are $357$ (fa+cn) WOS events among the $E_{582}$,
but STOA and STOA$^\prime$ only correctly predict $35\%$ of the WOS events, or $126$ 
events. STOASSN, STOASSN$^\prime$ and STOAAD perform much better for the WOS events, with correct 
nulls $235$, $235$ and $201$, respectively. 
The last column in table \ref{tbl:4result} shows the success 
rates of each models.
The sr of STOASSN and STOASSN$^\prime$ are  $65\%$, which are the highest among the four models, 
and the sr of STOA yields the lowest value of $0.51$. 

Table \ref{tbl:newdt} shows comparison of STOA and STOA$^\prime$ 
in terms of the forecast errors $\Delta T$. Among the $E_{582}$, there are $225$
WS events, for which the root mean square errors of $\Delta T$, 
$RMS_{\Delta T}(all)$, are listed in column 
2 of table \ref{tbl:newdt}. The $RMS_{\Delta T}(all)$ of STOA and STOA$^\prime$ are 
$18.96$ 
hours and $17.89$ hours, respectively. Column 3 shows the number of  events with 
$|\Delta T |\le 48$ hours. For STOA and STOA$^\prime$, $220$ and $222$ events are 
with $|\Delta T |\le 48$ hours, respectively. In addition, the root mean square 
errors of $|\Delta T |\le 48$ hours events are $17.21$ hours and $16.52$ hours for
STOA and STOA$^\prime$, respectively. It is shown that the new method STOA$^\prime$ 
not only provides prediction with more events with $|\Delta T |\le 48$ hours, but 
also gets smaller $RMS_{\Delta T}$. STOA$^\prime$ also performs better than STOA 
does with events with $|\Delta T |\le 24$ hours. 
 We also find that among the $225$ flare events 
accompanied with shocks at the Earth, STOA$^\prime$ provides better transit time for
 $124$ events. So statistically STOA$^\prime$ provides better forecast for the data 
set used in our study. 

Table \ref{tbl:newcriterion} shows the performances of the four 
different SATs prediction models: STOA, STOA$^\prime$, STOASSN 
and STOAAD, in terms of standard meteorological forecast skill scores. 
PODy implies a SATs model's performance with WS events that are 
predicted correctly. PODy of STOA equals $0.77$ means that STOA forecasts $77\%$ of 
the WS events successfully.  PODy of STOA$^\prime$ equals $78\%$, which 
is roughly the same as that of STOA.  But STOASSN, STOASSN$^\prime$ and STOAAD have less 
hits with WS events, so PODy of them are lower, which are $0.63$,
$0.64$ and $0.66$, respectively. 
Among the $E_{582}$, there are $357$ WOS events. So it is essential for an SATs
prediction model to provide satisfactory prediction for WOS events in practice. 
The second parameter PODn in table \ref{tbl:newcriterion} represents 
the proportion of WOS events that are predicted correctly. The models STOA and
STOA$^\prime$ still have the same performance in terms of PODn. 
 The PODn of other three models, STOASSN, STOASSN$^\prime$,
and STOAAD, are $0.66$, $0.66$ and $0.56$, 
respectively, which means that STOASSN and STOASSN$^\prime$ perform
 much better with WOS events than STOA. STOAAD performs not so well as STOASSN and STOASSN$^\prime$, but still it performs much better than STOA. 
Therefore,  FARs of the models with higher PODn, which is  the the proportion of wrong `yes' predictions, are kept smaller.
The smallest FAR is provided by STOASSN and STOASSN$^\prime$, and the largest one is provided by STOA and STOA$^\prime$. 
BIAS equals to `yes' predictions divided by `yes' observations, whose ideal value is
$1$.  The BIAS of STOA, STOA$^\prime$, STOASSN$^\prime$, STOASSN, and STOAAD are $1.80$, $1.80$, 
$1.17$, $1.18$, and $1.36$, respectively. So STOASSN yields the best BIAS.
The Critical Success Index CSI is used to evaluate  prediction 
of the events with low probability, which is the larger the better. 
It is shown that STOASSN and STOASSN$^\prime$ still yield the best CSI.  TSS, HSS and GSS, ranging in $[-1,1]$, $[-1,1]$ and $[-1/3,1]$, respectively, can 
be used to test whether the prediction is better than a random forecast. 
A positive value of TSS, HSS or GSS means that the prediction is better than random 
forecast, and the larger value is the better.  The scores of the models forecast
show that STOASSN and STOASSN$^\prime$ provide much better prediction of the $E_{582}$ than 
STOA does. STOAAD performs not so well as STOASSN and STOASSN$^\prime$, but it performs better than STOA.
The root mean square errors of $\Delta T$ for the `Hit' events listed in column 10 
of table \ref{tbl:newcriterion} show weak difference between the four models.
Finally, a $\chi^2$ test is used to  check the dependence between 
the observation and prediction. And the p-values show that 
we can have enough confidence in these four models.

Form table \ref{tbl:newcriterion} we can see that the models STOA and
STOA$^\prime$ yield the same standard meteorological forecast skill scores, except
that the $RMS_{\Delta T}$ of STOA$^\prime$ is smaller. In addition, STOA$^\prime$ 
is relatively simpler than STOA with a constant background plasma speed instead of
spacecraft measurement of solar wind speed at $1$ AU.

%----------------------------------------------------------------
\section{Conclusion and Discussion}
Prediction of the shock arrival times (SATs) at Earth is 
key to the space weather forecast, and 
many SATs prediction models are developed in the community. The three most
famous SATs prediction models STOA, ISPM and HAFv.2 usually all yield
prediction errors near $12$ hours. 
And among the three models, STOA is much easier to 
operate.   In this paper we make some improvements of 
STOA and develop four new methods STOA$^\prime$, STOASSN, STOASSN$^\prime$ and STOAAD.

STOA assumes the shock rides over a uniform background 
flow, the speed of which is very important for the
calculation of the shock's transit time. However it is 
impossible to measure the speed of plasma flow passed by the shock during all of its
 transition from Sun to Earth. STOA used the $1$ AU solar 
wind speed $V_{sw}$ at the time of parent flare event eruption 
as the background flow speed. However, in method STOA$^\prime$, a constant $V_c$ to
represent a typical solar wind speed, is used to calculate the transit time
of a shock from Sun to Earth. Our results show that the STOA$^\prime$ provides a 
better prediction than STOA. In space weather forecast practice, the STOA$^\prime$
is still effective when the solar wind speed observation is not available.
In addition, as a simpler method, the performance of STOA$^\prime$ is at least as 
good as that of STOA.   

STOA uses the simplified magnetoacoustic Mach number 
$M_{\alpha}$ to decide whether the shock will arrive 
at Earth. It is shown that STOA tends to yield too 
much `yes' prediction, which would undermine the continuity 
of the spacecraft operations in practice. In the method STOASSN, 
we develop a new criteria $M_{\alpha}^{SSN}$ by combing 
the effects of $M_{\alpha}$ and the sunspot number. We show that the method 
$M_{\alpha}^{SSN}$ improves the prediction of WOS flare events significantly. 

 We introduce a new model STOASSN$^\prime$ which combines the STOA$^\prime$ and STOASSN. It improves the prediction of the WOS events and the calculation of the transit time of the shocks.

The strength of the shock front will decrease with the increasing of the
distance from the shock nose. So it is more possible for a shock to be detected
when its nose moves towards the observer. The method STOAAD uses the angle  
distance between the shock nose direction and the 
Sun-Earth line to help deciding whether the shock 
will arrive at the Earth. The new criterion is very 
helpful to provide better prediction for WOS events.

In summary, all of the new models, STOA$^\prime$, STOASSN, STOASSN$^\prime$ and STOAAD make 
improvement of STOA. 
STOA$^\prime$ performs better on the calculation 
of transit time of shock than STOA does. However, more work should be done to 
further optimize the transit time calculation. The other three methods, 
STOASSN, STOASSN$^\prime$ and STOAAD, perform much better on the prediction of the WOS events 
than STOA, but they also miss more shocks than STOA. So further study are also 
needed to improve the criteria to predict the possibility of shock's  arrival.

 All the new models described in the paper are both developed and tested using 
the same events, $E_{582}$, which can be called all-event-models. It is more 
reasonable to build models using some sample
 events and test them using other events. So we choose one third events of $E_{582}$
as learning sample to develop new models which can be called learning-sample-models
(not shown here).
 To exclude the effects of solar cycle, we pick up every three events from the list.
The rest two thirds events are used to test the learning-sample-models. However, it 
can be shown that the learning-sample-models are roughly the same as the 
all-event-models in this work, so we only show the all-event-models.
%-----------------------------------------------------------------

%%------------------------------------------------------------------------ 

% \bibliography{reference.bib}

%%------------------------------------------------------------------------ %%
%
%  REFERENCE LIST AND TEXT CITATIONS
%
%% ------------------------------------------------------------------------ %%

%\noindent[width=20pc]
\clearpage
\clearpage

%\noindent[width=20pc]
\clearpage
\clearpage
%=======================================
\clearpage
\begin{table*}
%\tiny
 \caption{The Fearless Forecast Number (FF) for the events excluded from lists of \citet{FryEA03}, \citet{McKennaLawlorEA06} and \citet{SmithEA09b}\label{tbl:ffn}}
\begin{flushleft}
\begin{center}
\begin{tabular*}{\textwidth}{@{\extracolsep{\fill}}llllllllllllllll}
\tableline 
%\multicolumn{14}{c}{}\\ 
$1$ & $2$ & $3$ & $7$ & $9$ & $19$ & $20$ & $21$ & $38$ & $44$ & $45$ & $47$ & $54$ & $80$ & $114$ \\
$139$ & $257$ & $258$ & $322$ & $323$ & $325$ & $339$ & $374$
& $392$ & $445$ & $454$ & $458$& $470$ & $478$ & $489$ \\
$542$ & $588$ & $617$ & $618$ & $619$ & $470$ & $478$ & $489$ & $542$ & $588$ & $617$ & $618$ & $619$ & & \\ 
\tableline
\end{tabular*}
\end{center}
\end{flushleft}
\end{table*}
%-------------------------------------------------------------------
%\clearpage
%\begin{landscape}
%\begin{table*}
%\small
%\caption{\bf Definition of the statistic parameters used for evaluating SATs prediction models \tablenotemark{a}}
%\label{tbl:4resultdf}
%\begin{flushleft}
%\begin{center}
%\begin{tabular}{lp{9cm}} % 
%\tableline
% Hit (h)& Predict a shock and observe one at the Earth within $\pm 24$ hours\\ 
%Miss (m)&  Shock is observed but not predicted within $1~5$ days of the solar flare event or shock is predicted but not within $\pm 24$ hours \\ 
%False Alarm (fa) & Shock is predicted but not observed within 
%$\pm 24$ hours\\ 
%Correct Null (cn) & Shock is not predicted and no one is observed within $1-5$ days of the solar flare event\\
%Success Rate (sr) & (h+cn)/(h+m+fa+cn)\\  
%Probability of detection, yes (PODy) & h/(h+m) \\
%Probability of detection, no  (PODn) & cn/(fa+cn)\\
%False alarm ratio (FAR)& fa/(h+fa)\\
%Bias &(h+fa)/(h+m)\\
%Critical success index (CSI) & h/(h+fa+m)\\
%True skill score (TSS) & PODy+PODn-1 \\
%Heidke skill score (HSS)& (h+cn-C1)/(N-C1)\\
%Gilbert skill score (GSS)&(h-C2)/(h+fa+m-C2)\\
%\tableline
%\end{tabular}
%\tablenotetext{a}{\small where N=h+m+fa+cn, C2=(h+m)(h+fa)/(N), and C1=C2+(fa+cn)(m+cn)/N}
%\end{center}
%5\end{flushleft}
%\end{table*}
%\end{landscape}
%\clearpage

%------------------------------------------------------------------------
\clearpage
\begin{table*}
%\tiny
 \caption{Comparison of the results obtained using different
models for samples during solar cycle
$23$ (Hit window size $\pm 24$
Hours)\label{tbl:4result}}
%Indices\tablenotemark{a}}
%\begin{flushleft}
\begin{tabular*}{\textwidth}{@{\extracolsep{\fill}}cccccccc}
\tableline
Status & Number of Events & Model & (h) & (fa) & (cn) & (m)  & (sr)\\
\tableline
Cycle 23 & $582$ & 
    STOA    & $174$ & $231$ & $126$ & $51$ & $0.51$ \\ 
& & STOA$^\prime$ & $175$ & $231$ & $126$ & $50$ & $0.52$ \\
& & STOASSN   & $142$ & $122$ & $235$ & $83$ & $0.65$ \\
& & STOASSN$^\prime$ & $144$ & $122$ & $235$ & $81$ & $0.65$\\
& & STOAAD  & $149$ & $156$  & $201$ & $76$ & $0.60$ \\
\tableline
\end{tabular*}
%\end{flushleft}
%\tablenotetext{a}{High level of significance with $p<0.05$.}
%\tablenotetext{b}{Lesser, but still acceptable, level significance
%with $0.05<p<0.2$.}
\end{table*}
\clearpage

%-----------------------------------------------------------
\clearpage
\begin{landscape}
\begin{table*}
%\tiny
 \caption{Statistical comparison of the performances of STOA and 
STOA$^\prime$ using the $582$ events during 
solar cycle 23 \label{tbl:newdt}}
\begin{flushleft}
\begin{center}
\begin{tabular*}{\textwidth}{@{\extracolsep{\fill}}l|ccccc}
\tableline
& \multicolumn{5}{c}{} \\
%\cline{2-4} \cline{5-7} \cline{8-10}
& \small RMS$_{\Delta T}$ (all) & \small Count ($\le 48$hrs) & \small RMS$_{\Delta T}$ ($\le$48hrs) & \small Count ($\le$24hrs) & \small RMS$_{\Delta T}$ ($\le$24hrs)\\
\tableline
 STOA         & $18.96$ & $220$  & $17.21$ & $185$ & $11.64$ \\
STOA$^\prime$ & $17.89$ & $222$  & $16.52$ & $186$ & $11.42$ \\
%STOASA         & $0.37$ & $0.39$ & $0.41$ & $0.42$\\
\tableline
\end{tabular*}
\end{center}
\end{flushleft}
\end{table*}
\end{landscape}
\clearpage

%---------------------------------------------------------------------------
\clearpage
\begin{landscape}
\begin{table*}
\small
 \caption{Statistical comparison of the performances of STOA,
STOASEP, STOAF and STOASF in terms of standard meteorological
forecast skill scores using the $582$ events during 
solar cycle 23 }%\tablenotemark{a}
\label{tbl:newcriterion}
\begin{flushleft}
\begin{tabular*}{\textwidth}{@{\extracolsep{\fill}}l|ccccccccccc}
\tableline
%& \multicolumn{11}{c}{} \\
%\cline{2-4} \cline{5-7} \cline{8-10}
& PODy  & PODn & FAR & BIAS & CSI & TSS & HSS & GSS & RMS$_{\Delta T}$(Hit) & $\chi^2$ & p-value  \\
\tableline
 STOA & $0.77$ & $0.35$ & $0.57$ & $1.80$ & $0.38$ & $0.13$ & $0.11$ & $0.06$ & $11.65$ & $10.39$&$0.0013$\\
 STOA$^\prime$ & $0.78$ & $0.35$ & $0.57$ & $1.80$ & $0.38$ & $0.13$ & $0.11$ & $0.06$ & $11.48$  &$11.18$ &$0.00083$\\
STOASSN & $0.63$ & $0.66$ & $0.46$ & $1.17$ & $0.41$ & $0.29$ & $0.28$ & $0.16$ & $11.51$& $46.63$& $<0.0001$\\
STOASSN$^\prime$ & $0.64$ & $0.66$ & $0.46$ & $1.18$ & $0.41$ & $0.29$ & $0.29$ & $0.17$ & $11.37$&$49.47$&$<0.0001$\\
STOAAD &$0.66$& $0.56$ & $0.51$& $1.36$& $0.39$ &$0.23$ & $0.21$ & $0.12$& $11.54$  &$28.07$ & $<0.0001$\\
\tableline
\end{tabular*}
\tablenotetext{a}{p$<$0.05 implies a high level of significance.}
\end{flushleft}
\end{table*}
\end{landscape}
\clearpage
%% %========================================================================

%%% End of body of article:

%%%%%%%%%%%%%%%%%%%%%%%%%%%%%%%%
%% Optional Appendix goes here
%
%%%%%%%%%%%%%%%%%
% Geophysical Research Letters only allows an appendix without a letter.
%% You can get this result with 
%  \section*{Appendix} 
%  or 
%  \section*{Appendix: Title}
%%%%%%%%%%%%%%%%%
%
% \appendix resets counters and redefines section heads
% but doesn't print anything. 
% After typing  \appendix 
%
% \section{Here Is Appendix Title}
% will print 
% Appendix A: Here Is Appendix Title
%
% \section*{Appendix}
% will print 
% Appendix 
%
% \section*{Appendix: Here Is Appendix Title}
% will print 
% Appendix: Here Is Appendix Title 
%
% For only 1 appendix \appendix \section{Appendix} is preferred.
% which will print 
% Appendix A

%%%%%%%%%%%%%%%%%%%%%%%%%%%%%%%%%%%%%%%%%%%%%%%%%%%%%%%%%%%%%%%%
%
% Optional Glossary or Notation section, goes here
%
%%%%%%%%%%%%%%
% Glossary only allowed in Reviews of Geophysics
% \section*{Glossary}
% \paragraph{Term}
% Term Definition here
%
%%%%%%%%%%%%%%
% Notation -- End each entry with a period.
% \begin{notation}
% Term & definition.\\
% Second Term & second definition.
% \end{notation}
%%%%%%%%%%%%%%%%%%%%%%%%%%%%%%%%%%%%%%%%%%%%%%%%%%%%%%%%%%%%%%%%
%
%  ACKNOWLEDGMENTS

\begin{acknowledgments}
Our work are partly supported by grants
NNSFC 41304135, NNSFC 41374177, and NNSFC 41125016, the CMA grant GYHY201106011, and the Specialized Research Fund for State Key Laboratories of China.We thank \citet{FryEA03}, 
\citet{McKennaLawlorEA06} and 
\citet{SmithEA09b} for the events lists used in our work, and  we also thank http://sidc.oma.be/html/dailyssn.html for the daily sunspot number data.
\end{acknowledgments}

%% ------------------------------------------------------------------------ %%
%
%  REFERENCE LIST AND TEXT CITATIONS
%
% Either type in your references using
% \begin{thebibliography}{}
% \bibitem
% Text
% \end{thebibliography}
%
% Or, 
%
% If you use BiBTeX for your References, please produce your .bbl
% file and copy the contents into your paper here.
%
% Follow these steps:
% 1. Run LaTeX on your LaTeX file.
%
% 2. Run BiBTeX on your LaTeX file.
%
% 3. Open the new .bbl file containing the reference list and
%   copy all the contents into your LaTeX file here.
%
% 4. Comment out the old \bibliographystyle and \bibliography commands.
%
% 5. Run LaTeX on your new file before submitting.
%
% AGU does not want a .bib or a .bbl file, but asks that you
% copy in the contents of your .bbl file here.

%\begin{thebibliography}{}
%\bibitem{jskilby}J. S. Kilby,
%``Invention of the Integrated Circuit,'' {\it IEEE Trans. Electron Devices,}
%{\bf ED-23,} 648 (2008).
%\end{thebibliography}

%% ------------------------------------------------------------------------ %%
%
%  END ARTICLE
%
%% ------------------------------------------------------------------------ %%

\end{article}

%% Enter Figures and Tables at here:

% When submitting articles through the GEMS system:
% COMMENT OUT ANY COMMANDS THAT INCLUDE GRAPHICS.

% Figure captions go below this illustration; Table captions above tables

% ONE-COLUMN figure/table, including eps graphics
%
% \begin{figure}
% \noindent\includegraphics[width=20pc]{samplefigure.eps}
% \caption{Caption text here}
% \end{figure}
% \end{document}
%
% \begin{table}
% \caption{}
% \end{table}
%
% ---------------
% TWO-COLUMN figure/table
%
% \begin{figure*}
% \noindent\includegraphics[width=39pc]{samplefigure.eps}
% \caption{Caption text here}
% \end{figure*}
%
% \begin{table*}
% \caption{Caption text here}
% \end{table*}
%
% see below for how to make landscape figures or tables

%%% End the article here:

\end{document}